\documentclass[prb,twocolumn,showpacs]{revtex4}
\usepackage[latin1]{inputenc}
\usepackage{graphicx}
\usepackage{amssymb}
\usepackage{amsmath}
\usepackage{xspace}
\usepackage{dcolumn}
\usepackage{bm}
\usepackage{multirow}
\usepackage{color}
\usepackage[extra]{tipa}

\newfont{\tensy}{cmsy10}

\newcommand{\ie}[0]{i.e.\@\xspace}
\newcommand{\eg}[0]{e.g.\@\xspace}

\newcommand{\UP}[0]{\uparrow}
\newcommand{\DO}[0]{\downarrow}

\newcommand{\on}{\hat{n}}

\newcommand{\om}[0]{\omega}

\newcommand{\kF}{k_\text{F}}

\newcommand{\nag}{{\phantom{\dag}}}

\newcommand{\las}[0]{\langle}
\newcommand{\ras}[0]{\rangle}

\newcommand{\la}[0]{\left\las}
\newcommand{\ra}[0]{\right\ras}
\newcommand{\ket}[1]{\left|#1\ra}
\newcommand{\bra}[1]{\la#1\right|}

\begin{document}


\title{Excitation spectra and spin gap of the half-filled Holstein-Hubbard model}

\author{Martin Hohenadler}
\affiliation{\mbox{Institut f\"ur Theoretische Physik und Astrophysik,
    Universit\"at W\"urzburg, 97074 W\"urzburg, Germany}}

\author{Fakher F. Assaad}
\affiliation{\mbox{Institut f\"ur Theoretische Physik und Astrophysik,
    Universit\"at W\"urzburg, 97074 W\"urzburg, Germany}}

\begin{abstract}
  Single-- and two-particle excitation spectra of the one-dimensional,
  half-filled Holstein-Hubbard model are calculated using the continuous-time
  quantum Monte Carlo method. In the metallic phase, the results are
  consistent with a Luther-Emery liquid that has gapped spin and
  single-particle excitations but a gapless charge mode. However, given the
  initially exponential dependence of the spin gap on the backscattering matrix
  element, the numerical excitation spectra appear gapless in the
  weak-coupling regime, and therefore resemble those of a Luttinger
  liquid. The Mott phase has the expected charge gap and gapless spin
  excitations. The Peierls state shows a charge, spin and single-particle gap, a soft phonon mode,
  backfolded shadow bands and soliton excitations. Arguments and numerical
  evidence for the existence of a nonzero spin gap throughout the metallic
  phase are provided in terms of equal-time spin and charge correlation
  functions.
\end{abstract}

\date{\today}


\maketitle

\section{Introduction}\label{sec:intro}

The interplay of electron-electron interaction and the coupling of electrons
to the lattice is a fascinating topic in condensed matter physics, with
direct experimental relevance in compounds such as transition metal
dichalcogenides.\cite{Sipos08} Theoretically, this problem is particularly
difficult in the case of non-Fermi liquids, \eg, Luttinger liquids realized
in quasi-one-dimensional compounds. The difficulty is associated with the
competition between the instantaneous Coulomb repulsion and the retarded,
phonon-mediated attractive interaction.

The Holstein-Hubbard model is one of the most frequently studied models in
this context. It combines a local electron-electron repulsion (with strength
$U$), a nearest-neighbor hopping and a local density-displacement coupling
between electrons and phonons. Even though it neglects anharmonicity effects
and corrections due to incomplete screening,\cite{Ho.As.Fe.12} it is believed
to capture the key physics.  In one dimension, where powerful numerical and
analytical methods are available, existing work has mostly focused on the
bipolaron problem of two electrons, see Refs.~\onlinecite{PhysRevB.71.184309}
and~\onlinecite{0034-4885-72-6-066501} for an overview, and the half-filled
band.\cite{Guinea83,PhysRevB.31.6022,Stein97,ClHa05,hardikar:245103,FeWeHaWeBi03,0295-5075-84-5-57001,1742-6596-200-1-012031,NiZhWuLi05,MaToMa05,PhysRevB.84.165123,TeArAo05,PhysRevB.76.155114,PhysRevB.83.033104}
Some results for intermediate band fillings are also
available.\cite{TeArAo05,PhysRevB.76.155114,hardikar:245103,PhysRevB.84.085127,Re.Ya.Li.11}
For vanishing Hubbard repulsion, the Holstein-Hubbard model reduces to
Holstein's molecular crystal model \cite{Ho59a} which has been extensively
studied at half filling, see
Refs.~\onlinecite{Hirsch83a,JeZhWh99,PhysRevB.71.205113,tam:161103,Ho.As.Fe.12}
and references therein. The Holstein-Hubbard model may also be investigated
in the framework of dynamical mean-field
theory,\cite{MeHeBu02,Koller04,KoMeHe04,PhysRevLett.92.106401,Werner07,PhysRevB.81.235113}
although non-Fermi liquid physics and spatial correlations are not captured.

Interaction effects are dominant for a half-filled one-dimensional system.\cite{Giamarchi}
In the Hubbard model, a repulsive electron-electron interaction leads to a
Mott insulator with dominant spin-density-wave correlations but no long-range
order (a detailed definition of the various phases considered in the
following is given in Sec.~\ref{sec:results}; see also Table~\ref{tab:gaps}).
If the interaction is attractive, arising from electron-phonon
coupling in the limit of high phonon frequencies, backscattering opens a
spin gap, but the system remains metallic. Hirsch and Fradkin\cite{Hirsch83a}
argue that the $U(1)$ charge symmetry is broken at finite phonon frequencies,
so that long-range order can exist at $T=0$ as a result of the Peierls
instability.\cite{Peierls} Alternatively, such a transition to a state with
two electrons at every other site can result from a nearest-neighbor
electron-electron repulsion that emerges from electron-phonon interaction.
In the static limit of classical phonons, exactly solvable in mean-field
theory, an insulating Peierls state exists for any nonzero electron-phonon
coupling.  To fully understand the role of quantum lattice fluctuations (a
finite phonon frequency, as relevant for experiments) has turned out to be a
complex problem that still poses a number of open questions.

Whereas early numerical results for the half-filled Holstein model suggested
the existence of an insulating Peierls state for any nonzero
coupling,\cite{Hirsch83a} in agreement with
strong-coupling\cite{Hirsch83a,PhysRevB.10.1896} and renormalization group
arguments,\cite{PhysRevB.71.205113,Ba.Bo.07} subsequent and more accurate
treatments strongly suggest that lattice fluctuations destroy the
charge-ordered state below a critical coupling strength.\cite{JeZhWh99}
Turning to the Holstein-Hubbard model, a metallic phase of the Holstein model
should survive for small enough values of the electron repulsion, as
predicted previously.\cite{Guinea83,PhysRevB.67.081102} Early
numerical studies focused on large values of $U$ where the system is only
metallic at the single point of the Mott insulator to Peierls insulator
transition.\cite{Fe.Ka.Se.We.2003,FeWeHaWeBi03} Evidence for a metallic
phase, whose extent strongly depends on the phonon frequency, was reported
later.\cite{ClHa05,PhysRevB.76.155114,0295-5075-84-5-57001} Recently,
work aiming at characterizing this intermediate phase has initiated a debate
about the possibility of dominant superconducting
correlations,\cite{PhysRevB.76.155114,PhysRevB.84.165123} the existence of a
spin gap,\cite{1742-6596-200-1-012031} the validity of Luttinger liquid
theory for retarded interactions,\cite{PhysRevB.84.165123} and the existence
of a metallic phase.\cite{PhysRevB.71.205113,Ba.Bo.07,PhysRevB.84.165123}

The metallic phase is of particular interest in relation to superconductivity
in quasi-one-dimensional or higher dimensional systems. Even though the
continuous $U(1)$ gauge symmetry cannot be broken in one dimension, dominant
superconducting correlations imply a tendency toward superconductivity in
higher-dimensional settings.  Initial numerical results for the metallic
phase in the Holstein-Hubbard model indicated a Luttinger liquid parameter
$K_\rho>1$,\cite{ClHa05,hardikar:245103,0295-5075-84-5-57001,1742-6596-200-1-012031}
which suggests dominant pairing correlations. However, direct calculations of
correlators instead reveal that charge-density-wave correlations
dominate.\cite{TeArAo05,PhysRevB.76.155114,PhysRevB.84.165123} This
contradiction has been attributed to finite-size
effects\cite{hardikar:245103} and logarithmic
corrections\cite{PhysRevB.84.165123} due to the retarded nature of the
phonon-mediated interaction. Dominant pair correlations together with
$K_\rho>1$ have been reported at quarter filling,\cite{HoAs12} and degenerate
pairing and charge correlations exist in the half-filled lattice Fröhlich
model.\cite{Ho.As.Fe.12}

Another controversial issue is the existence of a spin gap in the metallic
phase, as a result of binding of electron pairs into singlet
bipolarons. This would imply that the low-energy theory is that of a {\it
  Luther-Emery liquid} (a Luttinger liquid with gapped spin
excitations but gapless charge excitations, see
Table~\ref{tab:gaps}),\cite{Lu.Em.74,Emery79} which has also been suggested
as the appropriate description of the normal state of Peierls compounds (\ie, for
$T>T_\text{c}$).\cite{Voit98} In the nonadiabatic limit, the mapping to the
attractive Hubbard model implies a nonzero spin gap. For small interactions,
the gap scales as\cite{Giamarchi} $\Delta_\sigma\sim
e^{-v_\text{F}/\mathcal{U}}$ where $\mathcal{U}$, the effective
backscattering matrix element, can either be related to the attractive
Hubbard interaction mediated by the phonons in the Holstein model, or to the
net interaction resulting from the interplay of electron-phonon and
electron-electron interaction in the Holstein-Hubbard model.  In the static
mean-field limit, a spin gap exists for any nonzero coupling. For finite and
especially for low phonon frequencies, there are significant retardation
effects. Similar to the destruction of the Peierls state by lattice
fluctuations due to a reduction of umklapp
scattering,\cite{JeZhWh99,Citro05,Giamarchi} the spin gap may be destroyed
for small enough interactions by a renormalization of backscattering under
the renormalization group flow.  Another possible mechanism is the
dissociation of bipolarons (a bound state always exists for exactly two
electrons, but this bipolaron is spatially extended for weak coupling) due to
mutual overlap, similar to polaron dissociation at finite band
filling.\cite{HoNevdLWeLoFe04,HoHaWeFe06} On the other hand, in the
low-energy limit, any finite phonon frequency may naively be argued to be
irrelevant, leading back to the spin-gapped attractive Hubbard model.\cite{Voit94}
There is also evidence from renormalization group calculations that
retardation can enhance backscattering.\cite{PhysRevB.71.205113} 

A gap that opens exponentially is of course very difficult to detect
by numerical methods. Several previous works explicitly state that the
metallic phase presumably has a finite spin
gap.\cite{hardikar:245103,PhysRevB.84.165123,1742-6596-200-1-012031} A
crossover from gapless to gapped spin excitations inside the metallic region
has also been proposed.\cite{0295-5075-84-5-57001,Assaad08} Recently, a spin
gap has been observed for all phonon frequencies in the quarter-filled
Holstein model at intermediate electron-phonon coupling.\cite{HoAs12}

Here, the continuous-time quantum Monte Carlo (CTQMC) method is used to
calculate the single-particle spectral function as well as the dynamical
charge and spin structure factors in all three phases of the Holstein-Hubbard
model. Previous work focused on single-particle
spectra\cite{FeWeHaWeBi03,NiZhWuLi05,MaToMa05} and the optical
conductivity.\cite{Fe.We.We.Go.Bu.Bi.2002,Fe.Ka.Se.We.2003} Additionally,
arguments and numerical evidence for the existence of a spin gap in the
entire metallic phase are presented. The paper is organized as follows: After
briefly introducing the model and the method in Secs.~\ref{sec:model}
and~\ref{sec:method}, the results are presented in Sec.~\ref{sec:results},
followed by the conclusions in Sec.~\ref{sec:conclusions}.

\section{Model}\label{sec:model}

In one dimension, the Holstein-Hubbard Hamiltonian can be written as
\begin{align}\label{eq:model}
  \hat{H}  
  &=
  -t\sum_{i \sigma} \left( c^{\dag}_{i\sigma} c^\nag_{i+1\sigma} + \text{H.c.} \right)
  + U \sum_i \on_{i\UP} \on_{i\DO}
  \\\nonumber
  &+ \sum_{i} \left( \frac{\hat{P}_{i}^2}{2M} + \frac{K \hat{Q}_{i}^2}{2}
     \right)
  - g \sum_{i} \hat{Q}_{i} \left( \hat{n}_{i}-1\right) 
  \,.
\end{align}
The first and second terms constitute the Hubbard model, describing electrons
with nearest-neighbor hopping $t$ and local repulsion $U$. The third and
fourth terms correspond to the lattice degrees of freedom and the
electron-phonon coupling. The phonons are described as harmonic oscillators
with frequency $\om_0=\sqrt{K/M}$, displacement $\hat{Q}_i$ and momentum
$\hat{P}_i$, and the coupling is of the density-displacement type proposed by
Holstein;\cite{Ho59a} $g$ is the coupling strength. The usual definitions
$\hat{n}_{i\sigma} = c^{\dag}_{i\sigma } c^\nag_{i\sigma }$, $\on_i =
\sum_\sigma \on_{i\sigma}$, and $n=\las \on_i\ras$ are used.

Using the path-integral representation, an effective, frequency-dependent
electron-electron interaction
\begin{equation}\label{eq:Uw}
  U(\om) = U + \lambda W \frac{\om_0^2}{\om^2-\om_0^2}
\end{equation}
can be derived, where $\lambda=g^2/(K W)$ is a dimensionless electron-phonon
coupling constant and $W=4t$ is the free bandwidth. In the nonadiabatic
limit (high phonon frequency, $\om_0\to\infty$) or at low energies
($\om\to0$), this interaction reduces to an instantaneous attractive or
repulsive Hubbard interaction $U_\infty=U-\lambda W$. In the following, the
hopping $t$ is taken as the unit of energy, and the lattice constant and
$\hbar$ are set equal to one.

\section{Method}\label{sec:method}

The model~(\ref{eq:model}) can be studied with the CTQMC method in the
interaction-expansion formulation.\cite{Rubtsov05} This method is free of Trotter
errors, and has been successfully applied to electron-phonon lattice
models.\cite{Assaad08,Hohenadler10a,Ho.As.Fe.12,HoAs12} The bosonic
degrees of freedom are integrated out exactly (without a cutoff for the
bosonic Hilbert space), and the resulting fermionic model with retarded
electron-electron  interactions is simulated.\cite{Assaad07} Methodological
details can be found in previous
publications\cite{Assaad07,Assaad08,Hohenadler10a,Ho.As.Fe.12,HoAs12} and a
review.\cite{Gull_rev} The numerical effort scales with the average expansion
order, making large system sizes accessible at weak coupling. The method
gives exact results with statistical errors for imaginary-time correlation functions.

The results below have been obtained at low but finite
temperatures $20<\beta t<162$; the inverse temperature is specified in each
figure caption. When comparing different system sizes, the ratio $\beta t/L$
has been kept constant. The system size was $L=30$ or $L=50$ for excitation
spectra, and $L=30$--162 for static correlation functions. Each QMC run was
started with a zero-vertex configuration, and equilibration was carried out
until the acceptance rates for addition and removal of a single
vertex have become equal to within one percent. Measurements were collected
from independent, parallel runs (typically 31) for each set of parameters.
For the most demanding parameters considered (Fig.~\ref{fig:dyn0.5peierls}), 
about 500 bins of 150 sweeps each were recorded (with measurements taken at
the end of each sweep). A sweep corresponded to 1000 proposed Monte Carlo
updates (addition or removal of a single vertex, or 32 attempted flips of
auxiliary Ising spins \cite{Assaad07}). Error analysis included the usual
binning and jackknife procedures to avoid underestimation of statistical
errors due to autocorrelations. Error bars shown indicate the standard error.

Spectral functions were obtained by applying a stochastic maximum-entropy
method\cite{Beach04a} to the imaginary-time Green function data and their covariance
matrix. Convergence of statistical errors was achieved over at least three
orders of magnitude. A flat default model was used,  and
the renormalization parameter was chosen to (approximately) achieve
$\chi^2=L_\tau$, where $L_\tau$ is the number of points on the imaginary time
axis.

The static correlation functions of interest are
\begin{align}
  S_\rho(r)   &=  \las (\on_r -n) (\on_0-n) \ras\,,\\\nonumber
  S^{zz}_\sigma(r) &= \las \hat{S}^z_r \hat{S}^z_0 \ras\,,\\\nonumber
  S^{xx}_\sigma(r) &= \las \hat{S}^x_r \hat{S}^x_0 \ras\,,\\\nonumber
  P(r)       &= \las \hat{\Delta}^\dag_r \hat{\Delta}^\nag_0 \ras \quad
  (\hat{\Delta}_r = c^\dag_{r\UP} c^\dag_{r\DO})\,,
\end{align}
measuring charge, spin, and s-wave pairing correlations, as well as their
Fourier transforms. In combination with the maximum entropy
method,\cite{Beach04a} excitation spectra can be calculated, including the
dynamical charge structure factor ($\Delta_{ji}=E_j-E_i$),
\begin{align}\label{eq:nqw}
  S_\rho(q,\om)
  &=
  \frac{1}{Z}\sum_{ij} {|\bra{i} \hat{\rho}_q \ket{j}|}^2
  e^{-\beta E_j} 
  \delta(\Delta_{ji}-\om)
  \,,
  \\
  \hat{\rho}_q 
  &= 
  \frac{1}{\sqrt{L}}
  \sum_r e^{iqr} (\on_{r} - n)\,,
\end{align}
the dynamical spin structure factor,
\begin{equation}\label{eq:sqw}
  S_\sigma(q,\om)
  =
  \frac{1}{Z}\sum_{ij} {|\bra{i} \hat{S}^z_q \ket{j}|}^2
  e^{-\beta E_j} 
  \delta(\Delta_{ji}-\om)
  \,,
\end{equation}
and the single-particle spectral function
\begin{align}\label{eq:akw}
  A(k,\om)
  =
  \frac{1}{Z}\sum_{ij\sigma}
  {|\bra{i} c_{k\sigma} \ket{j}|}^2 (e^{-\beta E_i}+e^{-\beta E_j})
  \delta(\Delta_{ji}-\om)
  \,.
\end{align}
Here $\ket{i}$ denotes an eigenstate with energy $E_i$.

\section{Results}\label{sec:results}

\begin{figure}
  \includegraphics[width=0.35\textwidth]{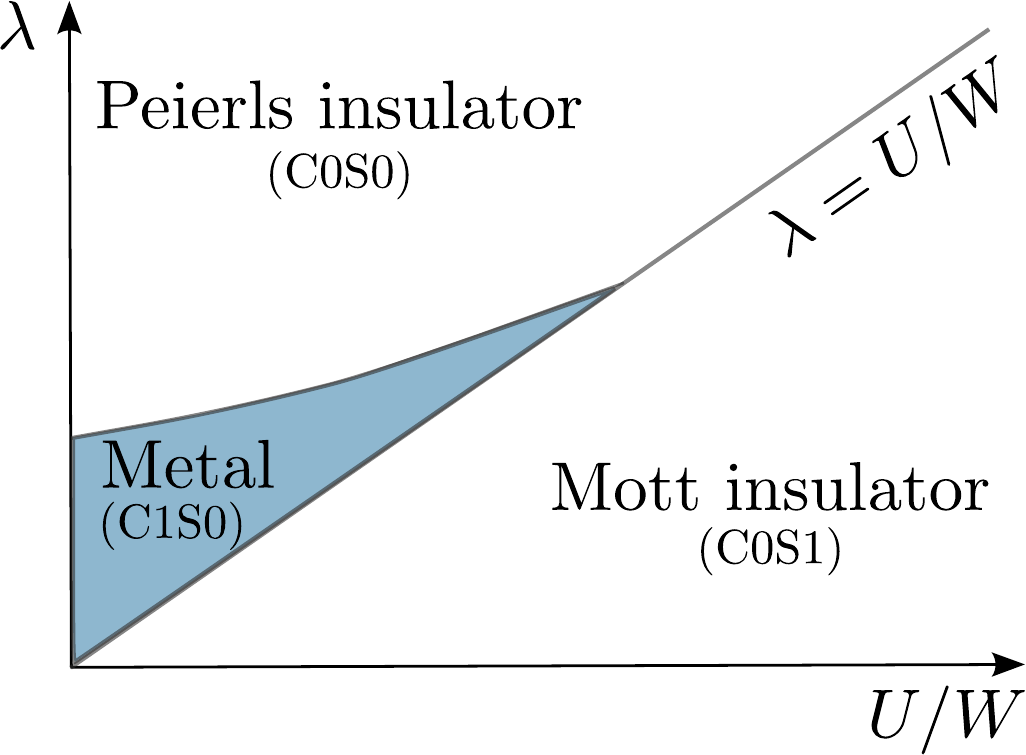}
  \caption{\label{fig:phasediagram} (Color online) 
    Schematic ground-state phase diagram of the half-filled, one-dimensional
    Holstein-Hubbard model with a finite phonon frequency, as suggested by
    numerical  simulations.\cite{ClHa05,hardikar:245103,PhysRevB.76.155114,0295-5075-84-5-57001,1742-6596-200-1-012031}
    For $U/W\gtrsim \lambda$, the system is a
    Mott insulator with dominant, power-law SDW correlations and a nonzero
    charge gap $\Delta_\rho$ (the notation C0S1 is explained in the text and
    in Table~\ref{tab:gaps}). For $\lambda\gg U/W$,
    the ground state is a Peierls insulator with {\it long-range} $2\kF$
    charge correlations and nonzero charge and spin gaps $\Delta_\rho$,
    $\Delta_\sigma$. For small $U/W$ and $\lambda$ (the scale depends on
    the phonon frequency), a metallic phase with gapless charge excitations
    exists for $\lambda\geq U/W$ ($U_\infty<0)$. Here it is argued that the
    entire metallic phase has a spin gap.
}
\end{figure}

Figure~\ref{fig:phasediagram} shows a schematic phase diagram of the
half-filled Holstein-Hubbard model. Exact numerical
investigations\cite{ClHa05,hardikar:245103,PhysRevB.76.155114,0295-5075-84-5-57001}
suggest the existence of three different phases at finite phonon frequencies.
These phases can be characterized and distinguished by the presence or
absence of an excitation gap for single-particle ($\Delta_\text{sp}$), charge
($\Delta_{\rho}$) and/or spin excitations ($\Delta_{\sigma}$) with zero momentum, see
Table~\ref{tab:gaps}. Following Balents and Fisher,\cite{PhysRevB.53.12133}
the notation C$x$S$y$ is used, where $x$ ($y$) is the number of gapless
charge (spin) modes (with $x,y=0,1$ for a strictly one-dimensional system).
Since a two-particle gap for either spin or charge excitations also
implies a nonzero single-particle gap $\Delta_\text{sp}$, knowledge of
$\Delta_{\rho}$ and $\Delta_{\sigma}$ is sufficient. Finally, because
low-energy charge transport is determined by long-wavelength density
fluctuations, the charge gap $\Delta_\rho$ further distinguishes metallic
($\Delta_{\rho}=0$) and insulating ($\Delta_{\rho}>0$) states.

The {\it Mott insulator} in Fig.~\ref{fig:phasediagram} exists for
repulsive interactions ($U_\infty\gtrsim 0$). It falls into the category C0S1,
with a nonzero charge gap $\Delta_\rho$ (and hence $\Delta_\text{sp}>0$)
reflecting the energy cost  for doubly occupied sites, a vanishing spin gap
$\Delta_\sigma=0$, and dominant, power-law spin-density-wave (SDW) correlations. 

The intermediate {\it metallic phase} exists for small enough $\lambda$ and $U$,
with the explicit scale depending on
$\om_0/t$.\cite{ClHa05,hardikar:245103,PhysRevB.76.155114,0295-5075-84-5-57001}
In one dimension, a metallic phase can either correspond to a {\it Luttinger
liquid} with $\Delta_\text{sp}=\Delta_\rho=\Delta_\sigma=0$ (C1S1), or to a {\it
Luther-Emery liquid} with $\Delta_\rho=0$ but
$\Delta_\text{sp},\,\Delta_\sigma>0$ (C1S0).\cite{Giamarchi}
The Luther-Emery liquid can be understood as a liquid of bosons, each
consisting of two electrons bound into a spin singlet. The excitation spectrum
for these bosons (corresponding to bipolarons in electron-phonon models) is
gapless within the metallic phase (hence $\Delta_\rho=0$), but the binding of
electrons into pairs gives rise to gaps for (electronic) single-particle and spin
excitations ($\Delta_\text{sp}\,,\Delta_\sigma>0$). The results presented below
suggest that in the Holstein and Holstein-Hubbard models, the metallic phase
is a Luther-Emery liquid (C1S0), except for the line $U_\infty=0$, which
belongs to the Luttinger liquid class C1S1.

\begin{table}[t]
  \caption{\label{tab:gaps}
  The different phases discussed in the text can be distinguished by the
  absence or presence of gaps for charge, spin, and single-particle
  excitations. The latter are denoted as $\Delta_\rho$, $\Delta_\sigma$, and
  $\Delta_\text{sp}$, respectively. The last column gives the corresponding
  class in the Balents-Fisher notation.\cite{PhysRevB.53.12133}
  }
  \begin{ruledtabular}
  \begin{tabular}{l|l|l|l|l}
    Phase                 & $\Delta_\rho$ & $\Delta_\sigma$ &
    $\Delta_\text{sp}$ & Class \\ \hline
    Luttinger liquid      &  zero        & zero          & zero & C1S1\\ 
    Luther-Emery liquid   &  zero        & nonzero       & nonzero & C1S0\\ 
    Mott insulator        &  nonzero     & zero          & nonzero & C0S1\\  
    Peierls insulator     &  nonzero     & nonzero       & nonzero & C0S0\\   
  \end{tabular}
  \end{ruledtabular}
\end{table}

The extent of the metallic phase increases with increasing phonon frequency.\cite{ClHa05,hardikar:245103,PhysRevB.76.155114,0295-5075-84-5-57001}
This can be ascribed to the suppression of charge order (the latter is
generally favored for $U_\infty<0$) due to enhanced lattice fluctuations. These fluctuations
have been shown\cite{JeZhWh99} to destroy the Peierls state of the
half-filled Holstein model below a critical value of $\lambda$. The numerical
results suggest that the metallic region exists in the Holstein-Hubbard model
for $U_\infty<0$, and is adiabatically connected to that of the Holstein
model ($U=0$). For $\om_0/t=5$, the density-matrix renormalization group
(DMRG)\cite{0295-5075-84-5-57001} yields a larger metallic region than the
QMC simulations,\cite{hardikar:245103} whereas for $\om_0/t=0.5$ the two
methods are in quite good agreement. These two frequencies will also be
considered here.  Finally, at the tricritical point where the Mott, Peierls
and metallic phases intersect, a first-order phase transition has been
reported.\cite{hardikar:245103}

When the electron-phonon interaction dominates ($U_\infty\ll0$), the system is a
{\it Peierls insulator} with nonzero single-particle, charge and spin gaps (C0S0).
The origin of the spin gap is again a pairing of electrons into spin
singlets, similar to the Luther-Emery phase. At $T=0$, the Peierls state has 
long-range charge-density-wave (CDW) order with period $q=2\kF$, corresponding
to a pair of electrons residing at every other lattice site. As in the
mean-field Peierls problem,\cite{Peierls,Hirsch83a} excitations out of the
energetically favored $2\kF$ state cost a finite energy $\Delta_\rho$.

\subsection{Spectra in the metallic phase}

\begin{figure}
  \includegraphics[width=0.4\textwidth]{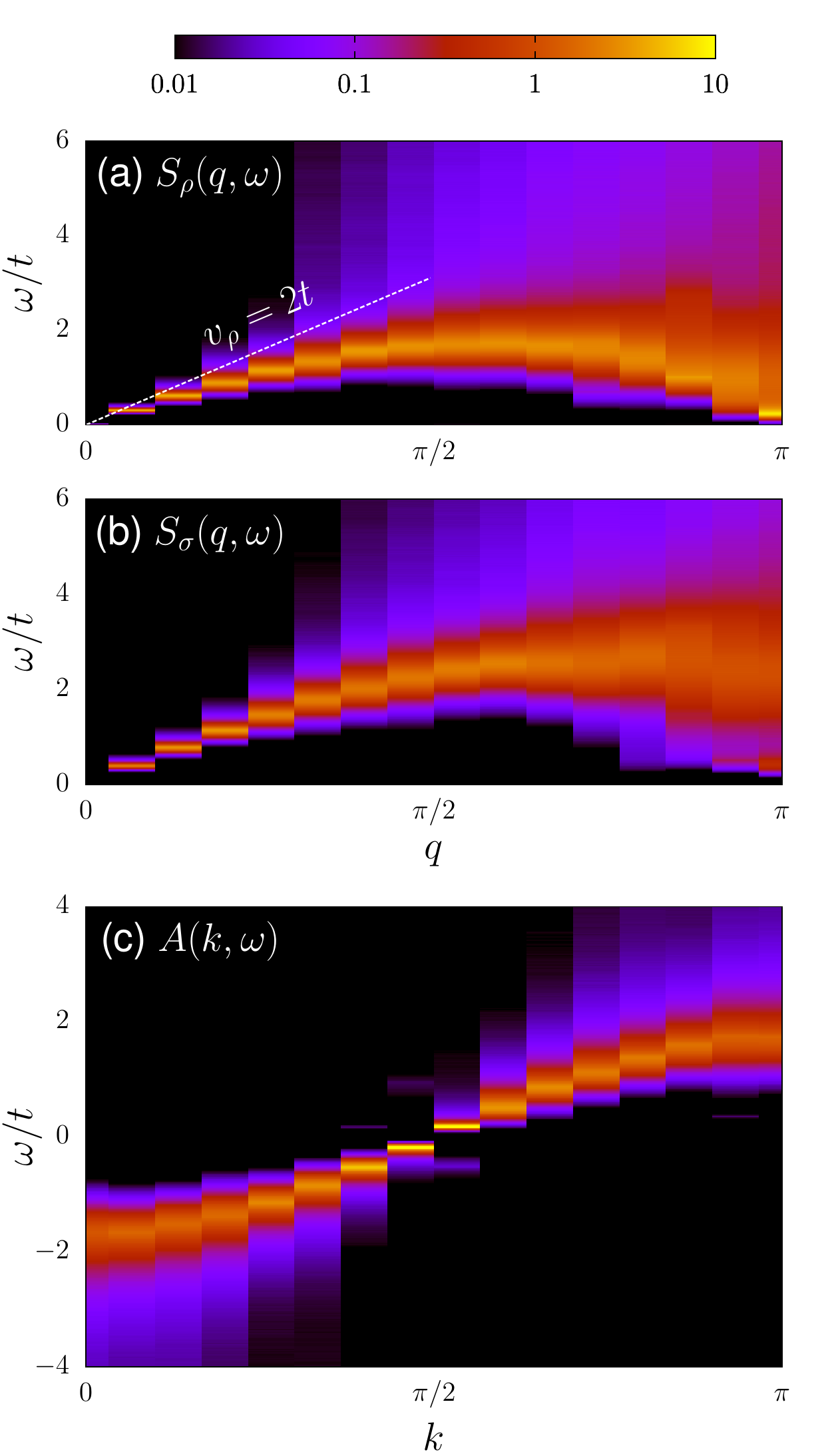}
  \caption{\label{fig:dyn5.0metal0.625} (Color online) 
    (a) Dynamical charge structure factor, (b)
    dynamical spin structure factor, and (c) single-particle spectral
    function for $\omega_0/t=5$, $U/t=1$, and $\lambda=0.5$, corresponding
    to the {\it metallic phase} (C1S0).\cite{hardikar:245103,0295-5075-84-5-57001} Here $L=30$,
    $\beta t=30$. The dashed line indicates the charge velocity in the
    noninteracting case ($\lambda=0$, $U=0$).
  }
\end{figure}

Figure~\ref{fig:dyn5.0metal0.625} shows the dynamical charge and spin
structure factors, defined in Eqs.~(\ref{eq:nqw}) and~(\ref{eq:sqw}), as well
as the single-particle spectral function [Eq.~(\ref{eq:akw})] in the
nonadiabatic regime $\om_0/t=5$ and for $U/t=1$, $\lambda=0.5$. For these
parameters, the metallic phase has been found to extend up to
$\lambda\approx0.65$ by QMC simulations,\cite{hardikar:245103} and up to
$\lambda\approx0.9$ by DMRG calculations.\cite{0295-5075-84-5-57001}

The dynamical charge structure factor, plotted in
Fig.~\ref{fig:dyn5.0metal0.625}(a), reveals a gapless mode at long
wavelengths. Its velocity, $v_\rho$, is noticeably smaller than the
noninteracting value $v_\rho=2t$ as a result of the enhanced mass of
bipolarons.  At $q=\pi=2\kF$, there is an almost completely softened
excitation with dominant spectral weight, which indicates strong but
power-law charge correlations with period $2\kF$ that are a precursor of the
ordered Peierls phase. The dynamical spin structure factor in
Fig.~\ref{fig:dyn5.0metal0.625}(b) also shows a linear mode. A possible spin
gap is too small to be visible for these parameters.

\begin{figure}
  \includegraphics[width=0.4\textwidth]{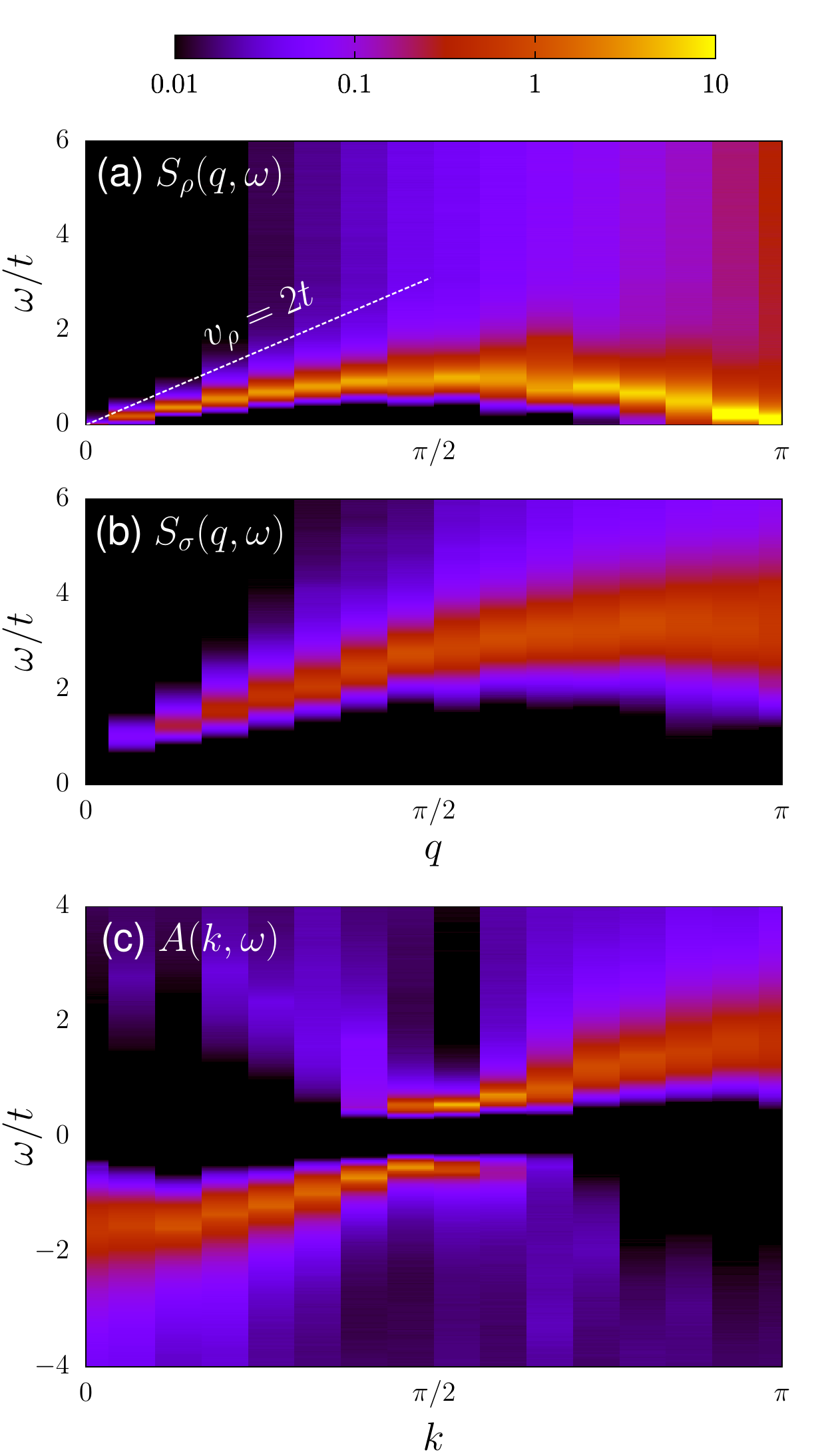}
  \caption{\label{fig:dyn5.0metal0.875} (Color online)    
    (a) Dynamical charge structure factor, (b)
    dynamical spin structure factor, and (c) single-particle spectral
    function for $\omega_0/t=5$, $U/t=1$, and $\lambda=0.875$, corresponding
    to the {\it metallic phase} (C1S0).\cite{0295-5075-84-5-57001} Here $L=30$,
    $\beta t=20$. The dashed line indicates the charge velocity in the
    noninteracting case ($\lambda=0$, $U=0$).
}
\end{figure}

\begin{figure}
  \includegraphics[width=0.4\textwidth]{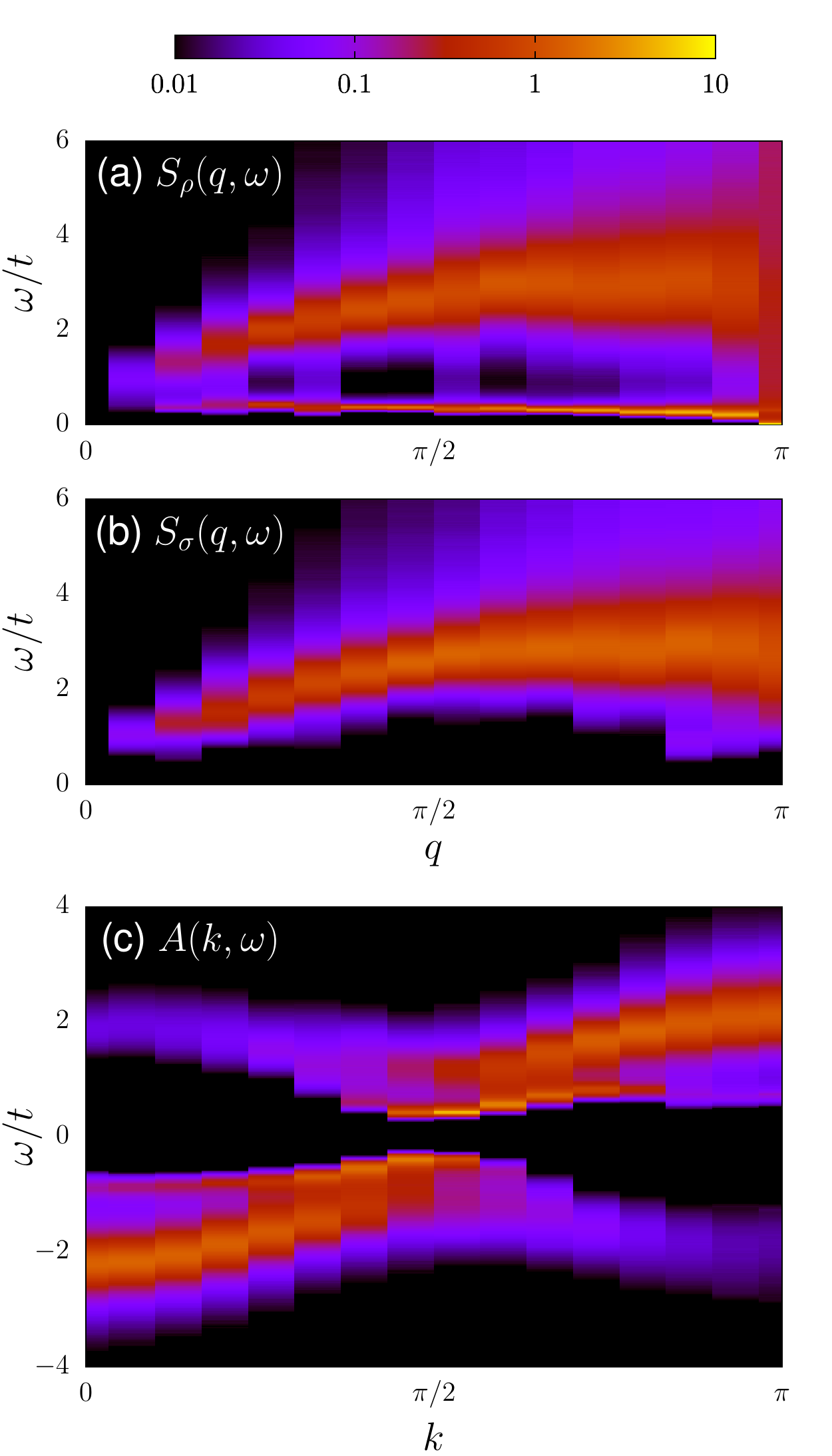}
  \caption{\label{fig:dyn0.5peierls} (Color online) 
   (a) Dynamical charge structure factor, (b)
    dynamical spin structure factor, and (c) single-particle spectral
    function for $\omega_0/t=0.5$, $U/t=0.2$, and $\lambda=0.4$, corresponding
    to the {\it Peierls insulator} (C0S0). Here $L=30$,
    $\beta t=30$.
}
\end{figure}

Figure~\ref{fig:dyn5.0metal0.875} shows the excitation spectra for a stronger
electron-phonon coupling $\lambda=0.875$, closer to (potentially on) the DMRG
phase boundary for the metallic phase.\cite{0295-5075-84-5-57001} The charge
structure factor still appears gapless, signaling metallic behavior, with an
even smaller charge velocity and stronger softening at $q=2\kF$ than in
Fig.~\ref{fig:dyn5.0metal0.625}(a). The spin structure factor has developed a
well visible gap at long wavelengths, which is also reflected in the
corresponding single-particle spectrum in
Fig.~\ref{fig:dyn5.0metal0.875}(c). The existence of bound singlets, as
reflected by the spin gap, makes bipolarons the low-energy degrees of
freedom.  The corresponding Drude weight strongly depends on the phonon
frequency.\cite{HoAs12} An important point concerning
Figs.~\ref{fig:dyn5.0metal0.625} and~\ref{fig:dyn5.0metal0.875} is that a
gapless mode with significant Drude weight exists in $S_\rho(q,\om)$ near
$q=0$. In contrast, the spectrum in the insulating Peierls phase [see
Fig.~\ref{fig:dyn0.5peierls}(a)] exhibits a strong depletion of spectral
weight and a finite gap for long-wavelength charge excitations.

The metallic region shrinks with decreasing phonon
frequency.\cite{0295-5075-84-5-57001,hardikar:245103,PhysRevB.76.155114} To
study the adiabatic regime, a value $\omega_0/t=0.5$ is considered, for which
the extent of the metallic phase is
known.\cite{0295-5075-84-5-57001,hardikar:245103} For $U/t=0.2$, it exists up
to $\lambda\approx0.25$.\cite{0295-5075-84-5-57001} Since the spin gap
increases on approaching the Peierls phase (see discussion above for
$\om_0/t=5$), an electron-phonon coupling close to the phase boundary
($\lambda=0.225$) is chosen.

\begin{figure}
  \includegraphics[width=0.4\textwidth]{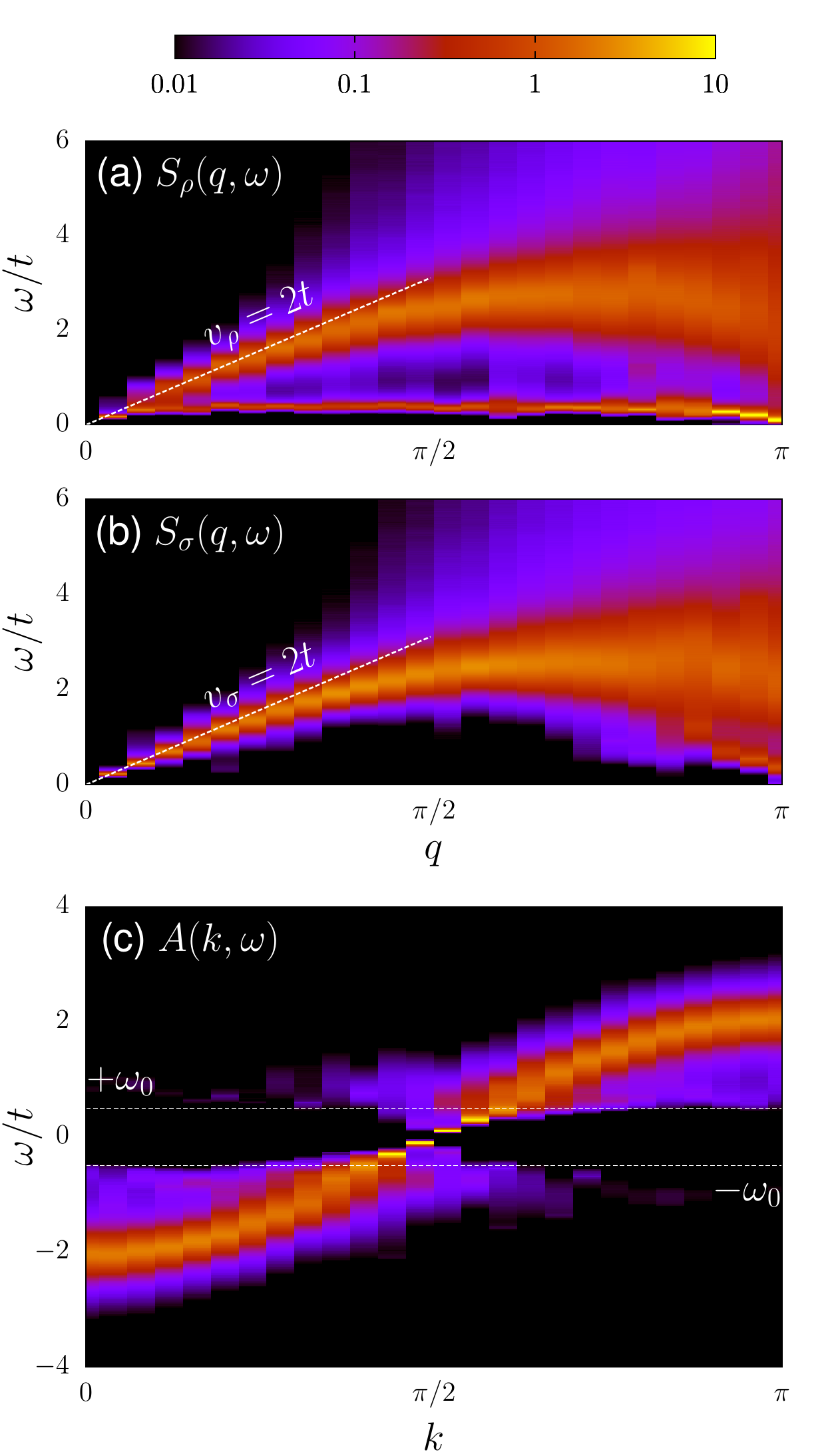}
  \caption{\label{fig:dyn0.5metal} (Color online)  
    (a) Dynamical charge structure factor, (b)
    dynamical spin structure factor, and (c) single-particle spectral
    function for $\omega_0/t=0.5$, $U/t=0.2$, and $\lambda=0.225$, corresponding
    to the {\it metallic phase} (C1S0). Here $L=50$,
    $\beta t=50$.  The dashed lines in (a) and (b) indicate the charge and
    spin velocities in the noninteracting case ($\lambda=0$, $U=0$),
    respectively. The horizontal lines in (c) mark $\pm\om_0$.
  }
\end{figure}

The charge velocity $v_\rho$ in Fig.~\ref{fig:dyn0.5metal}(a) is almost
unchanged compared to the noninteracting case, which can be understood in
terms of the rather small bare coupling constant $g$. Hence, the electrons
are only loosely bound into large bipolarons. An important difference to the
case $\omega_0/t=5$ considered before is that the charge structure factor in
Fig.~\ref{fig:dyn0.5metal}(a) reveals the renormalized phonon frequency at
low energies. The latter is partially softened at $q=2\kF$, which is a
previously noted precursor effect of the Peierls
transition.\cite{Hohenadler06,CrSaCa05,SyHuBeWeFe04,Hohenadler10a}

The spin structure factor in Fig.~\ref{fig:dyn0.5metal}(b) appears to be
gapless (although it be will argued below that a finite spin gap exists
throughout the metallic phase), with a weakly renormalized, linear low-energy
mode. Finally, the single-particle spectrum in Fig.~\ref{fig:dyn0.5metal}(c)
also appears gapless, with small but visible signatures of the hybridized
polaron modes located near $\om=\pm\om_0$.\cite{Assaad08}

\begin{figure}
  \includegraphics[width=0.45\textwidth]{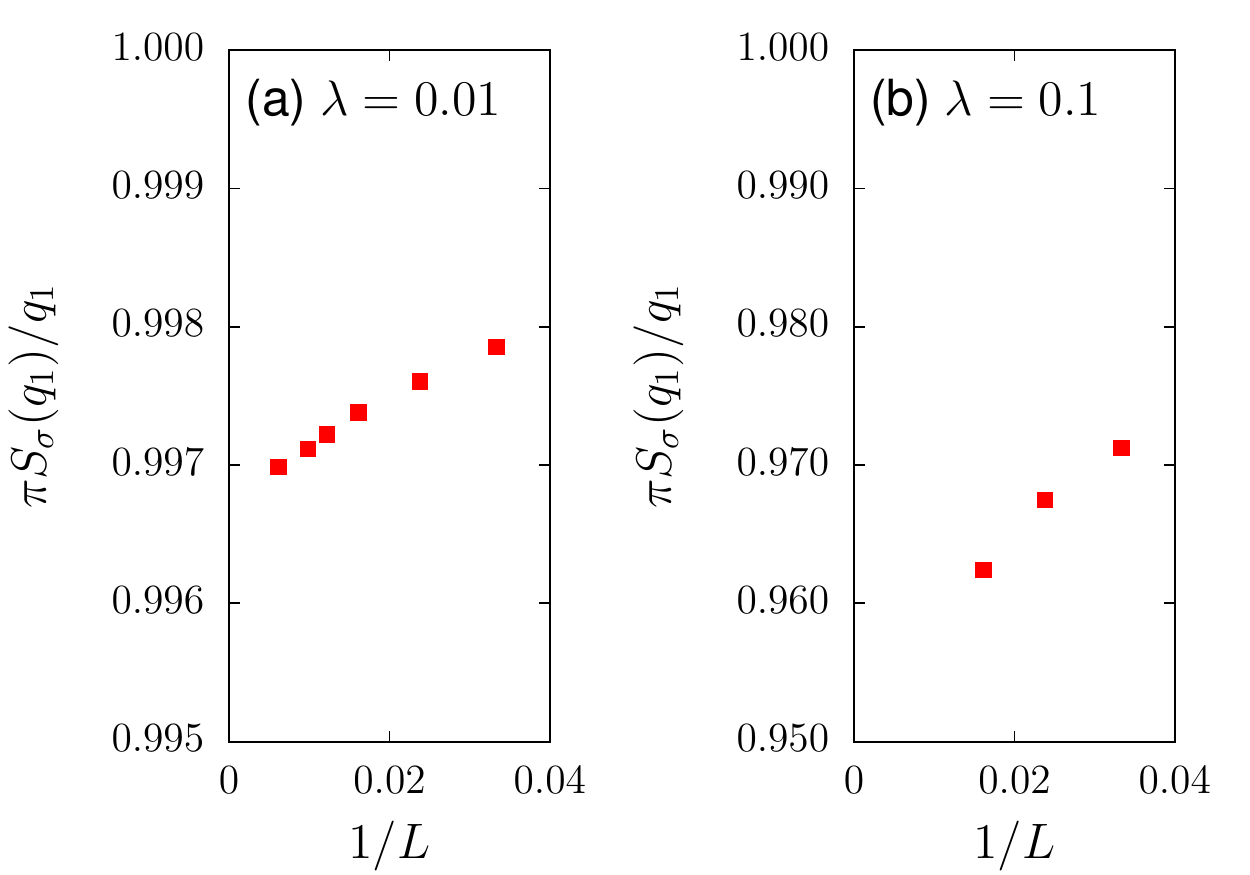}
  \caption{\label{fig:Ksigma} (Color online) 
  Renormalized spin structure factor $\pi S_\sigma(q_1)/q_1$ at $q_1=2\pi/L$,
  related to  $K_\sigma$ via
  $K_\sigma=\lim_{L\to\infty} \pi S_\sigma(q_1)/q_1$, as
  a function of inverse system size in the metallic phase of the Holstein model.
  Results are for $\beta t=L$, $\om_0/t=0.5$, $U=0$, and (a) $\lambda=0.01$,
  (b) $\lambda=0.1$.
}
\end{figure}

\begin{figure*}
  \includegraphics[width=\textwidth]{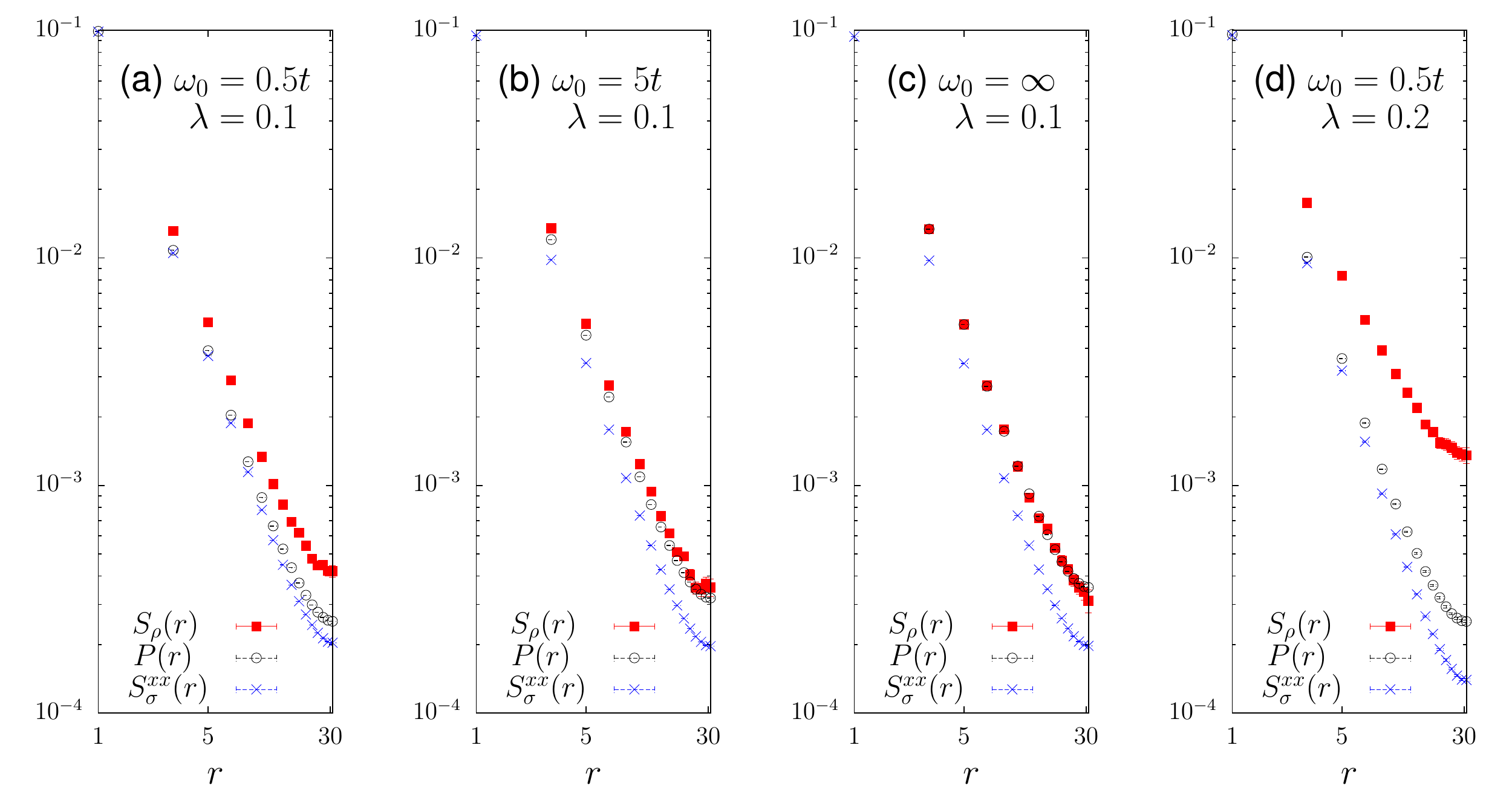}
  \caption{\label{fig:correlations} (Color online) 
  Real-space charge, pairing, and spin correlations (all rescaled by a factor
  $1/2$) for (a) $U=0$, $\lambda=0.1$, $\om_0=0.5t$, (b) $U=0$, $\lambda=0.1$,
  $\om_0=5t$, both corresponding to the metallic phase of the Holstein model. 
  (c) The same correlation functions for the attractive Hubbard model
  with $U=-0.4t$. (d) As in (a), but for $U=0$, $\om_0=0.5t$, and
  $\lambda=0.2$. Results are for $\beta t = L = 62$. Spin correlations in
  the $z$ and $x$ directions are identical within error bars; $S^{xx}_\sigma$
  is shown because of its smaller error bars.
}
\end{figure*}

Previous numerical calculations of excitation spectra of the half-filled,
metallic Holstein-Hubbard model were restricted to the single-particle
spectrum,\cite{NiZhWuLi05} and showed in particular that spin-charge
separation is only visible for large values of $\lambda$ (this in turn
requires a large ratio $\om_0/t$ for a metallic state to exist), in
accordance with simulations at quarter filling.\cite{Assaad08} This
conclusion explains the absence of such features in the present results. The
spectrum was also calculated at $U=0$.\cite{Ho.As.Fe.12} Analytical results
include the exact spectrum of a Luttinger liquid (\ie, without a spin gap)
coupled to phonons,\cite{Meden94} as well as approximate results for the
spectral function of the Luther-Emery liquid.\cite{Voit98} The latter work
predicts that branch cuts may be regularized by the existence of a gap.
Interestingly, for small $\lambda$ where the spin gap is not resolved and
where spin-charge separation is not detected in numerical simulations, the
spectrum is very well described by analytical results for a spinless
Luttinger liquid coupled to phonons,\cite{Meden94} and closely resembles the
spectrum of the spinless Holstein model.\cite{Hohenadler10a}

\subsection{Spin gap in the metallic phase}

A spin gap is visible in the dynamical correlation functions shown above, at
least close to the Peierls phase boundary. The initially exponential dependence of this
gap on the interaction strength and restrictions in system size make it
difficult to detect the spin gap at weak coupling by considering excitation
spectra. Instead, evidence for a nonzero spin gap in the metallic
Holstein-Hubbard model comes from numerical results for the Luttinger liquid
parameter $K_\sigma$, which can in principle be determined from the spin
structure factor and finite-size extrapolation as $K_\sigma=\lim_{L\to\infty}
\pi S_\sigma(q_1)/q_1$, with $q_1=2\pi/L$. For the half-filled Hubbard model,
were complications such as retarded electron-electron interactions are
absent, numerical simulations generically give $\pi S_\sigma(q_1)/q_1>1$ for
$U>0$, $\pi S_\sigma(q_1)/q_1<1$ for $U<0$, and $\pi S_\sigma(q_1)/q_1=1$
exactly at $U=0$.\cite{hardikar:245103} In particular, although logarithmic
corrections make it hard to demonstrate $K_\sigma=1$ or 0 in the spin gapless
and spin gapped phases, respectively, the finite-size estimates always
decrease with increasing system size.  Therefore, for the repulsive case
$U>0$, $\pi S_\sigma(q_1)/q_1$ slowly approaches the value 1 implied by spin
rotation symmetry from above, whereas for the attractive case $U<0$, $\pi
S_\sigma(q_1)/q_1$ deviates more and more from 1.  Within a low-energy
theory, and given spin symmetry, the latter result can only be understood in
terms of $K_\sigma=0$ and a spin gap.  Similar behavior can also be observed
beyond the Hubbard model, and values $\pi S_\sigma(q_1)/q_1<1$ have been
argued to represent empirical evidence for a spin gap.\cite{PhysRevB.59.4665}
For a more detailed discussion see
Refs.~\onlinecite{PhysRevB.59.4665,hardikar:245103}.

Figure~\ref{fig:Ksigma} shows the system size dependence of $\pi
S_\sigma(q_1)/q_1$ in the adiabatic regime ($\om_0/t=0.5$) of the Holstein
model in the metallic phase. Even for very weak coupling $\lambda=0.01$, $\pi
S_\sigma(q_1)/q_1<1$ and monotonically decreasing with increasing system
size. At stronger coupling $\lambda=0.1$, the size dependence is
significantly more pronounced. A similar picture persists for larger phonon
frequencies.  The spin gap has also been measured directly by means of the
DMRG method.  While $\Delta_\sigma$ is clearly finite for selected parameters
in the metallic region, it is difficult to decide if this is true of the
whole metallic phase.\cite{0295-5075-84-5-57001,1742-6596-200-1-012031} In
particular, the exponential dependence of the gap on the coupling makes it
practically impossible to detect the critical point directly from
$\Delta_\sigma$.

To provide further evidence for the existence of a spin gap, it is useful to
consider the real-space correlation functions. In a gapless Luttinger liquid,
the slowest decaying correlations are given by \cite{Voit94}
\begin{align}\label{eq:correl}
  R_\text{CDW}(r) &\sim \cos (2\kF r) r^{-K_\rho-K_\sigma}\,,\\\nonumber
  R^{zz}_\text{SDW}(r) &\sim \cos (2\kF r) r^{-K_\rho-K_\sigma}\,,\\\nonumber
  R^{xy}_\text{SDW}(r) &\sim \cos (2\kF r) r^{-K_\rho-1/K_\sigma}\,,\\\nonumber
  R_\text{SS}(r) &\sim r^{-1/K_\rho-K_\sigma}\,.
\end{align}
For a generic spin-rotation invariant system with $K_\sigma=1$, spin and
charge correlations are degenerate, and dominate over pairing correlations if
$K_\rho<1$. On the other hand, attractive interactions ($K_\rho<1$) lead to
dominant pairing correlations. Additional complications are logarithmic
corrections arising from marginally relevant operators. In the repulsive
Hubbard model, such corrections cause spin correlations to dominate over
charge correlations.\cite{Schulz90} As pointed out by Voit,\cite{Voit98}
dominant $q=2\kF$ CDW correlations cannot occur in a gapless Luttinger
liquid. (Although logarithmic corrections could in principle favor charge
over spin correlations, there is no known example of such behavior. Moreover,
such corrections would not explain $K_\sigma<1$.)

In contrast, due to the exponential suppression of spin correlations, a
Luttinger liquid with a gap for spin excitations supports dominant $2\kF$
charge correlations. In terms of the low-energy model, this case arises when
attractive backscattering is taken into account.\cite{Lu.Em.74} The
corresponding model is often referred to as a Luther-Emery model. A familiar
example is the attractive Hubbard model. That attractive backscattering can
originate from electron-phonon interaction is illustrated by the previously
mentioned explicit relation between the Holstein model and the attractive
Hubbard model.  To obtain the correlation functions of the a Luther-Emery
liquid, the spin gap can formally be accounted for by setting $K_\sigma=0$,
leading to\cite{Voit94} $R_\text{CDW}(r) \sim \cos(2\kF r) r^{-K_\rho}$ and
$R_\text{SS}(r) \sim r^{-1/K_\rho}$, see also Eq.~(\ref{eq:correl}). Hence,
neglecting possible logarithmic corrections, repulsive (attractive)
interactions lead to dominant charge (pairing) correlations.

Results for the charge, spin and pairing correlation functions are shown in
Fig.~\ref{fig:correlations}. Figures~\ref{fig:correlations}(a) and (b)
correspond to the adiabatic ($\om_0/t=0.5$) and the nonadiabatic regime
($\om_0/t=5$), respectively.  To be able to explore very weak interactions
$U_\infty$, it is advantageous to consider $U=0$, so that the metallic phase
exists down to $\lambda=0$. This choice also eliminates any uncertainties
about the phase boundaries of the metallic region.  For the coupling
$\lambda=0.1$ chosen, charge correlations dominate over pairing and spin
correlations, as expected for a Luther-Emery liquid with repulsive
interactions. This dominance is more pronounced for $\om_0/t=0.5$
[Fig.~\ref{fig:correlations}(a)] than for $\om_0/t=5$
[Fig.~\ref{fig:correlations}(b)] because the system is closer to the Peierls
phase, and cannot be explained by a rescaling of the constant (\ie,
independent of $r$) prefactors of $2\kF$ correlations in the charge and spin
channels. The results are qualitatively similar also at weaker interactions
(the values investigated where as small as $\lambda=0.01$), but it becomes
increasingly more difficult to distinguish the different correlation
functions on finite systems.

For comparison, Fig.~\ref{fig:correlations}(c) presents the same correlation
functions for the attractive Hubbard model with $U=-\lambda W=-0.4t$,
corresponding to the limit $\om_0=\infty$ at the same coupling $\lambda=0.1$
used in Figs.~\ref{fig:correlations}(a), (b). Whereas spin correlations are
again clearly suppressed, the mapping between the attractive and the
repulsive Hubbard model at half filling implies a degeneracy of charge and
pairing correlations that is captured, within statistical errors, by the
numerical data. Finally, Fig.~\ref{fig:correlations}(d) shows results at a
stronger coupling $\lambda=0.2$, for which charge correlations and the
suppression of spin correlations are much more pronounced. The exponential
decay of spin correlations has been observed before in DMRG
calculations.\cite{PhysRevB.76.155114}

The results for $\pi S_\sigma(q_1)/q_1$ and the real-space correlation
functions are fully compatible with the existence of a spin gap throughout
the metallic phase.  In particular, the behavior of these observables is
essentially identical to the attractive Hubbard model, for which the
existence of a spin gap is well established. In contrast to the gapless
Luttinger liquid model, the Luther-Emery model hence provides a consistent
description of the numerical data. The existence of a spin gap for any 
nonzero electron-phonon coupling in the Holstein model has also been
predicted based on renormalization group calculations.\cite{PhysRevB.54.2410,PhysRevB.71.205113}

The above results suggest that the half-filled, metallic Holstein model
behaves as a repulsive Luther-Emery liquid, with dominant charge
correlations, and a nonzero spin gap in the thermodynamic limit that
suppresses spin correlations with respect to charge and pairing correlations.
Assuming that the metallic phase is the same at $U=0$ and $U>0$ leads to the
conclusion that the whole metallic region of the Holstein-Hubbard model is a
Luther-Emery liquid. The exponential length scale related to the spin gap is
expected to lead to a crossover (as a function of system size) from the
correlation functions~(\ref{eq:correl}), as appropriate for a system with
gapless spin excitations, to the corresponding analytical results for the
Luther-Emery liquid.\cite{Voit94} In particular, the relation
$\alpha_\text{CDW}\alpha_\text{SS}=1$, expected for a Luther-Emery liquid, is
not obeyed by the numerically determined power-law exponents. Whereas this
fact has previously been attributed to retardation
effects,\cite{PhysRevB.84.165123} it can be expected to be a general problem
associated with finite-size simulations of Luther-Emery liquids.

\subsection{Spectra in the Peierls phase}

The single-particle spectral function in the Peierls phase has been
calculated before using numerical methods.\cite{FeWeHaWeBi03,Ho.As.Fe.12}
Bosonization results for the spectrum of a quarter-filled CDW insulator are
also available.\cite{Essler.Tsvelik.2002} The dynamical charge structure
factor has been studied for the spinless Holstein model.\cite{Hohenadler10a}
Using the CTQMC method, the Peierls state is most accessible in the adiabatic
regime. Specifically, $\om_0/t=0.5$, $\lambda=0.4$, and $U/t=0.2$ are
considered, and results are shown in Fig.~\ref{fig:dyn0.5peierls}.

The dynamical charge structure factor in Fig.~\ref{fig:dyn0.5peierls}(a) is
characterized by a finite gap and strongly suppressed spectral weight for
long-wavelength charge excitations, and by a soft phonon mode with gapless
excitations at $q=2\kF$. Similar to the metallic Luther-Emery phase, the
Peierls state has a finite (but larger) spin gap, see
Fig.~\ref{fig:dyn0.5peierls}(b).  Compared to the spin structure factor in
the metallic phase, Fig.~\ref{fig:dyn0.5metal}(b), there is also a finite gap
for spin excitations with $q=2\kF$.

The single-particle spectrum in Fig.~\ref{fig:dyn0.5peierls}(c) reveals the
expected Peierls gap, as well as clear signatures of backfolded shadow bands
at high energies---as a result of
dimerization\cite{Vo.Pe.Zw.Be.Ma.Gr.Ho.Gr.00}---and soliton excitations at
low energies. The latter have previously been observed in the spinless
Holstein model,\cite{Hohenadler10a} and the extended Holstein model with
nonlocal interactions,\cite{Ho.As.Fe.12} and distinguish the otherwise
qualitatively (apart from the size of the single-particle gap) similar
spectra of the metallic phase [Fig.~\ref{fig:dyn0.5metal}(c)] and the Peierls
phase [Fig.~\ref{fig:dyn0.5peierls}(c)].

\begin{figure}
  \includegraphics[width=0.4\textwidth]{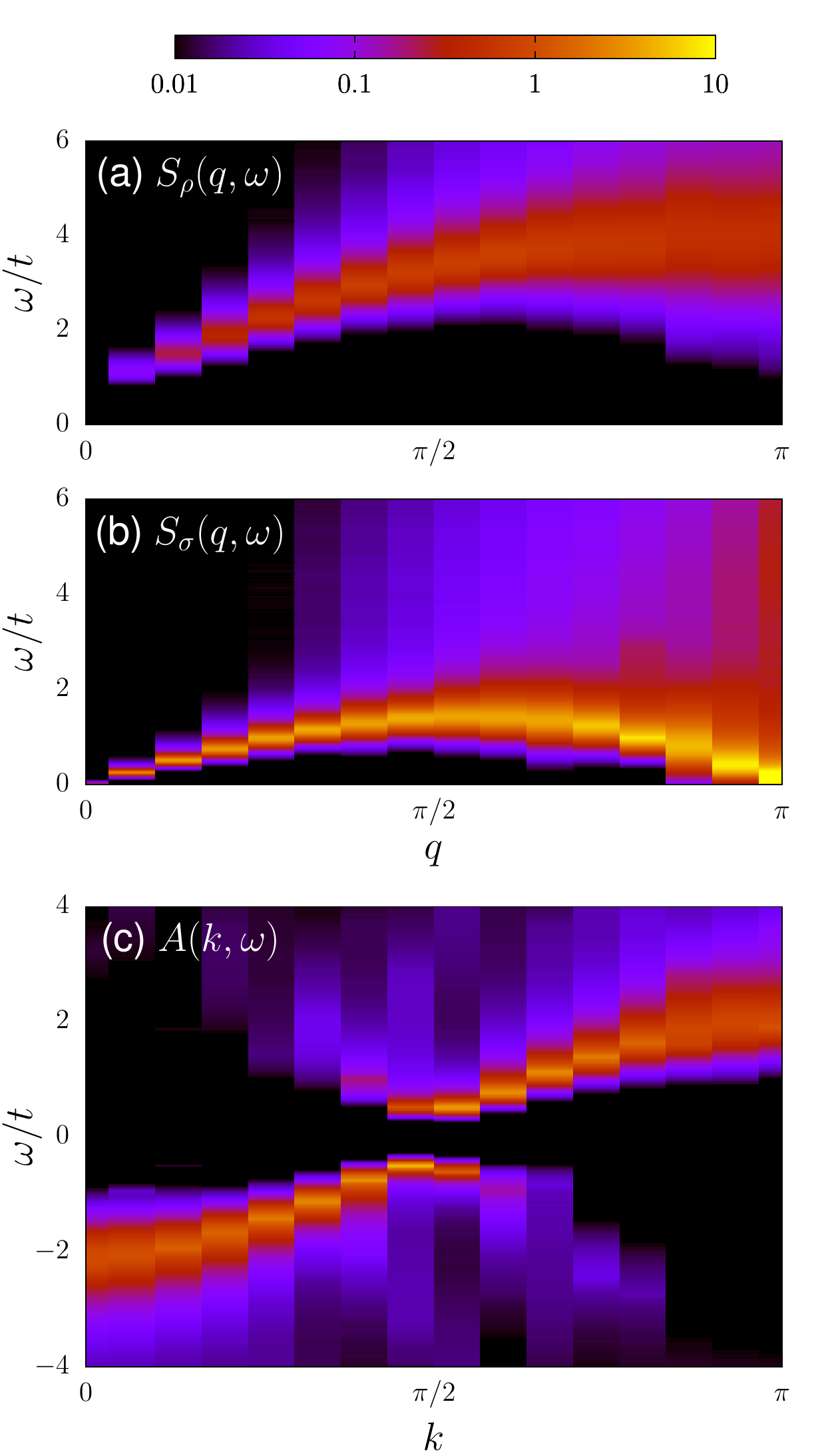}
  \caption{\label{fig:dyn5.0mott} (Color online) 
    (a) Dynamical charge structure factor, (b)
    dynamical spin structure factor, and (c) single-particle spectral
    function for $\omega_0/t=5$, $U/t=4$, and $\lambda=0.25$, corresponding
    to the {\it Mott insulator} (C0S1). Here $L=30$,
    $\beta t=20$.
}
\end{figure}

\subsection{Spectra in the Mott phase}

The excitation spectra in the Mott phase are shown in
Fig.~\ref{fig:dyn5.0mott}, for the parameters $\om_0/t=5$, $\lambda=0.25$,
and $U/t=4$. The single-particle spectrum in this phase has previously been
calculated.\cite{FeWeHaWeBi03,NiZhWuLi05,MaToMa05}

In accordance with the discussion of the phase diagram at the beginning of
Sec.~\ref{sec:results}, the results in Fig.~\ref{fig:dyn5.0mott} reveal
gapped charge excitations but gapless spin excitations, \ie, just opposite to
the metallic phase shown in Fig.~\ref{fig:dyn5.0metal0.875}. Although the
spin symmetry is not broken in the Mott phase, the spin structure factor
reveals gapless spin excitations at $q=2\kF$ reminiscent of the soft phonon
mode related to long-range charge order in
Fig.~\ref{fig:dyn0.5peierls}(a). For the value $\om_0/t=5$ considered, the
renormalized phonon mode is not visible in $S_\rho(q,\om)$, and it is
strongly suppressed compared to the Peierls state even for $\om_0/t=0.5$
(data not shown).

The single-particle spectrum in Fig.~\ref{fig:dyn5.0mott}(c) bears a close
resemblance to the results for the Hubbard model with the same
$U/t=4$,\cite{PhysRevB.66.075129} although signatures of spin-charge
separation are suppressed here as a result of a reduced, effective
interaction ($U_\infty/t=3$) and thermal fluctuations at $T>0$.  The effects
of the latter are quite subtle for intermediate values of
$U/t$.\cite{PhysRevB.66.075129} Instead of the pronounced backfolded shadow
bands related to long-range charge order in the Peierls state,
Fig.~\ref{fig:dyn0.5peierls}(c), there is rather incoherent spectral weight
with no clear dispersion. Additionally, the hallmark soliton excitations
visible in Fig.~\ref{fig:dyn0.5peierls}(c) are completely absent.

\section{Conclusions}\label{sec:conclusions}

The single-- and two-particle excitation spectra of the one-dimensional,
half-filled Holstein-Hubbard model have been calculated with the
continuous-time quantum Monte Carlo method. The spectra in the metallic phase
are consistent with a Luther-Emery liquid that has gapless charge excitations
but gapped spin excitations with a gap that opens exponentially as a function
of the interaction strength. In the Peierls phase, the spectra reveal both a charge and a
spin gap, and a soft charge mode at $q=2\kF$. The single-particle spectral
function reveals backfolded shadow bands and soliton excitations. The Mott
phase has a charge gap but no spin gap, as well as a soft spin mode despite
the absence of long-range order. No clear signatures of spin-charge
separation were observed.

The static charge and spin correlation functions have been calculated on
rather large systems and reveal dominant $2\kF$ charge correlations even for
very weak coupling. Because such behavior cannot occur in a generic Luttinger
liquid, this observation can be regarded as evidence for a nonzero spin
gap.\cite{Voit94} In particular, the behavior of these correlation functions
is essentially the same as for the attractive Hubbard model (for which the
existence of a spin gap is generally accepted) even for low phonon
frequencies. When combined with the fact that the Luttinger parameter
$K_\sigma$ is found to be less than unity even for very weak coupling, the
numerical data are naturally explained by assuming that the metallic phase of
the Holstein-Hubbard model is a Luther-Emery liquid, with electrons paired
into singlet bipolarons. Earlier reports of both gapless and gapped metallic
regions\cite{0295-5075-84-5-57001,Assaad08} can be attributed to the
exponentially small size of the gap which makes its detection very
demanding. The existence of a nonzero spin gap also has important
implications for the low-energy description of this model. In particular, the
spin gap is not captured by the exact results available for the spectral
function of the Luttinger model with phonons.\cite{Meden94} Finally, given
the relevance of backscattering at any band filling, the Luther-Emery
description can be expected to be generic for metallic phases in
Holstein-type models, in accordance with the suggestive physical picture of
electron pairs that are bound into bipolarons.

{\begin{acknowledgments}%
    Computer time at the J\"ulich Supercomputing Centre and financial support
    from the DFG (Grant No.~Ho~4489/2-1) are gratefully acknowledged, as are
    discussions with S. Ejima, H. Fehske, V. Meden, A. Sandvik, D. Schuricht,
    S. Wessel, and, in particular, F. Essler.
\end{acknowledgments}}


\end{document}